# Engineering Magnetization with Photons: Nanoscale Advances in the Inverse Faraday Effect for Metallic and Plasmonic Systems

Chantal Hareau[1], Xingyu Yang[1], Maria Sanz[1], Matthew Sheldon[2,3] and Mathieu Mivelle[1] *

[1]Sorbonne Université, CNRS, Institut des NanoSciences de Paris, INSP, F-75005 Paris, France
[2]Department of Chemistry, University of California, Irvine, Irvine, CA 92697, USA.
[3]Department of Materials Science and Engineering, University of California, Irvine, Irvine, CA 92697, United States

*Corresponding authors: mathieu.mivelle@sorbonne-universite.fr

The inverse Faraday effect, the ability of light to act as a source of magnetism, is a cornerstone of modern ultrafast optics. Harnessing this effect at the nanoscale promises to transform data storage and spintronics, yet its predictive understanding remains elusive. This review synthesizes recent progress in engineering the IFE within plasmonic architectures. We bridge the theoretical foundations, from classical drift-current models to quantum descriptions, with the latest experimental milestones, including pump-probe studies that have verified the effect's sub-picosecond nature. Special emphasis is placed on how nanostructure design allows for unprecedented control, enabling functionalities like chiral or reversed magnetization by locally sculpting the optical spin density.

Despite this progress, a crucial challenge pervades the field: a stark, often orders-of-magnitude, mismatch between predicted and measured magnetization values. We contend that resolving this discrepancy is paramount. The path forward requires the development of novel experimental probes capable of directly imaging these fleeting magnetic fields at their native length and time scales, ultimately unlocking the true potential of nanoscale optical magnetism.

## 1. An Introduction to the Inverse Faraday Effect

Over the past few decades, the field of light-induced magnetism has emerged and grown significantly, primarily driven by the potential impact of controlling magnetic order at ultrafast timescales—an exciting prospect for both fundamental research and data storage technologies.[1–4] This field was initiated by the pioneering work of Beaurepaire *et al.*, who studied ultrafast demagnetization using pulsed lasers.[5] In 2005, Kimel *et al.* demonstrated that spin dynamics can be manipulated using a femtosecond laser pulse with circular polarization, through a helicity-dependent and non-thermal process.[6] Building on these findings, Stanciu *et al.* showed that magnetic domains could be deterministically reversed in ferrimagnetic materials using only circularly polarized light,[7] and Lambert *et al.* later extended this observation to ferromagnets, suggesting that the underlying all-optical helicity-dependent switching mechanism arises from a fairly general light–matter interaction.[8] To date, two promising (yet still debated) candidates have been proposed to explain these phenomena: the inverse Faraday effect and magnetic circular dichroism.[2,9]

Leveraging these pioneering contributions, a multitude of studies and research topics have surfaced, all aiming to control magnetization at increasingly smaller spatial scales and ever faster temporal scales.[2–4,10] One of these areas focuses on generating magnetization in non-magnetic metallic materials — particularly plasmonic metals — through a near-instantaneous, nonlinear process that imparts continuous circular motion to the electrons, thereby producing a magnetic field.[11–13] This process underpins the inverse Faraday effect in non-magnetic metals.

The aim of this review is to present both theoretical frameworks and recent experimental results related to the inverse Faraday effect in non-magnetic metals. We begin by outlining the various theoretical descriptions, with particular emphasis on the simpler and more intuitive classical approach. We then discuss both numerical and experimental investigations on this topic, especially those involving plasmonic metals at the nanoscale, highlighting robust findings as well as discrepancies among different studies. In the discussion section, we identify persisting gaps in our understanding of the Inverse Faraday Effect (IFE) in metals and propose several key questions for future research, emphasizing the fascinating opportunities these developments offer to the magnetics community.

The inverse Faraday effect describes a light–matter interaction in which a medium becomes magnetized by circularly polarized light (Figure 1).[14,15] The induced magnetization follows the light's propagation direction, and its sign is determined by the light's helicity. In particular, right-handed circularly polarized light magnetizes the medium along the propagation direction, while left-handed circularly polarized light magnetizes it in the opposite direction. This effect was first predicted and observed in the 1960s, with an initial phenomenological description framed in analogy to the conventional Faraday effect.[14,15] As the potential key to unlocking the all-optical control of magnetism, the inverse Faraday effect offers a fertile ground for new scientific breakthroughs and technological applications.

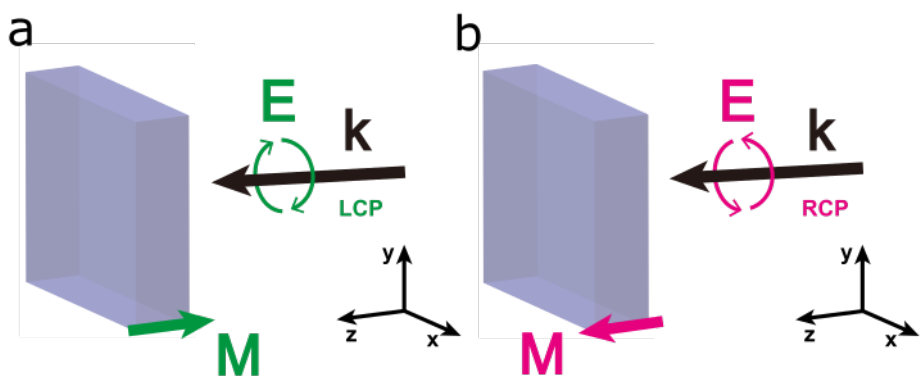

*Figure 1: Illustration of the inverse Faraday effect. When light is left circularly polarized, the induced magnetization is oriented in the direction opposite to the wave propagation (a). Conversely, when light is right circularly polarized, the magnetization is oriented along the direction of propagation (b). Figure adapted from [16].*

**Towards nanoplasmonics**

Nanoplasmonics is a subfield of plasmonics that explores the interactions between light and metallic nanostructures at sub-diffraction scales. Its primary focus lies in the excitation of localized surface plasmon resonances (LSPRs), which are collective oscillations of conduction electrons driven by incident light. These resonances yield strong electromagnetic field confinement and significant field enhancement near the nanostructures.

Beyond its fundamental interest, nanoplasmonics enables practical applications by taking advantage of pronounced local field enhancements. For instance, it facilitates surface-enhanced Raman scattering (SERS),[17,18] boosts fluorescence signals,[19–21] and supports nonlinear optical effects. As a result, nanoplasmonics underpins cutting-edge developments in advanced sensing, imaging, and photonic device fabrication.[22–25] In addition, researchers can tailor the optical response of photonic nanostructures—such as nanoparticles, nanorings, and nanoantennas—by precisely controlling parameters like size, shape, and material composition. This versatility is paving the way for innovations in biosensing, photovoltaics, and quantum information processing.[26–29]

One particularly promising direction is the integration of nanoplasmonics and magneto-optics, which opens new avenues in photonics and spintronics. By merging plasmonic nanostructures with magnetic materials, researchers exploit localized electromagnetic fields to significantly amplify magneto-optical phenomena, including Faraday rotation, Kerr effects, and magnetic circular dichroism.[3,4,30,31] Such enhancements hold potential for ultrafast optical switching, high-density data storage, and sensitive magneto-optical sensors. Notably, the tunable nature of plasmonic resonances allows devices to operate across a broad wavelength range—an essential feature for integrated optical circuits and quantum information technologies.

The ability of plasmonic nanostructures to concentrate light energy at subwavelength scales makes them integral to investigations of ultrafast magneto-optics. Both intrinsically plasmonic metals and hybrid heterostructures combining plasmonic and magnetic materials have been studied.[32–38] A consensus emerging from these efforts is that plasmonic resonances can enhance optically induced magnetization—alongside other magneto-optical effects.

Recently, the IFE in non-magnetic plasmonic metals has garnered attention. Compared to magnetically ordered materials—which involve additional complexities arising from spin interactions—non-magnetic metals offer a simpler platform for exploring this phenomenon. Theoretical work suggests that these metals can maintain, or even exceed, the relevant performance metrics achieved by magnetic materials in terms of field intensity, spatial confinement, and ultrafast dynamics.[31,39–42] Indeed, plasmonic nanostructures can spatially confine induced magnetic fields at the nanoscale—below the diffraction limit—and generate them on ultrafast timescales using femtosecond laser pulses, a tool now widely employed in advanced photonic systems.

**Classical Framework for the Inverse Faraday Effect in Metals**
A simplified classical theory of IFE in metals was described by Hertel in 2006,[11] and further developed by others in the following years.[12,13,43,44] It starts simply by analyzing the forces that light applies onto the free electrons in a collisionless plasma, which have an effect on circulating electron currents.

To clarify the theoretical framework underlying the inverse Faraday effect, we begin with the continuity equation, which describes the conservation of electron density $n$ in space and time:

$$\frac{\partial n}{\partial t} + \vec{\nabla} \cdot (n\vec{v}) = 0 \qquad (1)$$

Multiplying by the elementary charge *e*, this equation represents charge conservation explicitly:

$$e\frac{\partial n}{\partial t} + \vec{\nabla} \cdot \vec{J} = 0 \qquad (2)$$

where $\vec{J} = en\vec{v}$ is the total conduction current density.

To analyze optical phenomena driven by an oscillating electromagnetic field (frequency *ω*), we separate the electron density *n* into a stationary average ⟨*n*⟩ and a small time-dependent fluctuation *δn*,

$$n = \langle n \rangle + \delta n \qquad (3)$$

with *δn* ≪ ⟨*n*⟩. This allows us to rewrite the continuity equation explicitly in terms of the fluctuating part:

$$e\frac{\partial(\delta n)}{\partial t} + \vec{\nabla} \cdot e(\langle n \rangle + \delta n)\vec{v} = 0 \qquad (4)$$

In linear optics, the dominant conduction current density $J_\omega$ is approximated by neglecting small density fluctuations, thus

$$\vec{J_\omega} \approx e\langle n \rangle \vec{v} \approx \sigma \vec{E} \qquad (5)$$

$\sigma$ being the dynamic conductivity and $\vec{E}$ the optical electric field.

Switching to frequency-domain analysis (assuming fields vary as $e^{-j\omega t}$, j being the imaginary unity), the continuity equation simplifies to:

$$e(-j\omega)\delta n + \vec{\nabla} \cdot \vec{J_\omega} = 0 \qquad (6)$$

yielding a direct expression for the fluctuating charge density:

$$\delta n = \frac{1}{j\omega e} \vec{\nabla} \cdot \vec{J_\omega} \qquad (7)$$

We now consider the total current density $\vec{J}$ explicitly, including second-order nonlinear terms arising from fluctuations:
It is this fluctuating part of the charge density (δn) that plays a crucial role in the theory of the IFE. Hence, in the more general expression for the total current density

$$\vec{J} = e(\langle n \rangle + \delta n)\vec{v} = \vec{J_\omega} + e\delta n\vec{v} \qquad (8)$$

Substituting the earlier expressions and using the linear relationship $\vec{v} = \frac{\sigma}{e\langle n \rangle}\vec{E}$, we obtain:

$$e(\delta n)\vec{v} = e(\frac{1}{j\omega e}\vec{\nabla}\cdot\vec{J_\omega})\vec{v} = e(\frac{1}{j\omega e}\vec{\nabla}\cdot\vec{J_\omega})(\frac{\sigma}{e\langle n\rangle}\vec{E}) \qquad (9)$$

Since physical measurable quantities are real and time-averaged, we take the time-average explicitly. Using standard complex notation for harmonic fields, the time-averaged product of fluctuating fields becomes:

$$\langle \vec{J}\rangle = e\langle \delta n \cdot \vec{v}\rangle = \frac{e}{4}(\delta n \cdot \vec{v}^* + \delta n^* \cdot \vec{v}) = -\frac{j}{4e\langle n\rangle\omega}\left[\vec{J_\omega}^*(\vec{\nabla}\cdot\vec{J_\omega}) - c.c.\right] \qquad (10)$$

where "c.c." indicates the complex conjugate of the preceding term. This result clearly isolates a second order nonlinear effect: the DC drift current $\vec{J_d} = e\langle \delta n \cdot \vec{v}\rangle$

Through a mathematical transformation for arbitrary vectors $\vec{A}$ and $\vec{B}$,

$$\vec{\nabla}\times(\vec{A}\times\vec{B}) = (\vec{B}\cdot\vec{\nabla})\vec{A} - \vec{B}(\vec{\nabla}\cdot\vec{A}) + \vec{A}(\vec{\nabla}\cdot\vec{B}) - (\vec{A}\cdot\vec{\nabla})\vec{B} \qquad (11)$$

The drift current in Eq. (10) can be separated into two components.

$$\vec{J_d} = -\frac{j}{4e\langle n\rangle\omega}\vec{\nabla}\times(\sigma^*\vec{E}^*\times\sigma\vec{E}) + \frac{1}{4e\langle n\rangle\omega}[j(\sigma^*\vec{E}^*\cdot\vec{\nabla})\sigma\vec{E} + c.c.] \qquad (12)$$

These two contributions represent the magnetization currents and the ponderomotive currents, respectively. Thus, the drift current can be simply expressed as:

$$\vec{J_d} = \vec{\nabla}\times\vec{M} + \vec{\Gamma} \qquad (13)$$

Where:
- $\vec{M}$ is the magnetization,
- $\vec{\nabla}\times\vec{M}$. represents the magnetization currents, and
- $\vec{\Gamma}$ represents the ponderomotive currents.

We now have a general equation that accounts for all the contributions to the inverse Faraday effect in metallic nanostructures. In this expression, $\vec{\nabla}\times\vec{M}$. and $\vec{\Gamma}$ represent the macroscopic contributions responsible for the collective and steady motion of the metal's electrons (a direct current (DC)), while $\vec{M}$ represents the microscopic contribution linked to the circular motion of electrons around their center of mass.

This expression of $\vec{M}$ is consistent with the phenomenological definition of IFE which states that the magnetization appears under circular (or elliptical) polarization of light only, and is linear with respect to the light intensity.

A key advantage of this classical theory is its straightforward implementation in commonly used finite-element simulations within the photonics community. By describing the Faraday effect in nano- and microscale systems without resorting to atomically resolved calculations, it provides a practical tool for modeling.[39,45] Additional work by other researchers has refined this classical approach by incorporating system dissipation and resonance effects.[41,43,46] Their intuitively derived equations predict an enhancement of the effect near the plasmon resonance. In particular, Battiato *et al.* emphasize that in dissipative media, the Faraday and inverse Faraday effects cannot always be described by the same Verdet constant—especially near a resonance—unlike what was assumed in earlier phenomenological descriptions.[43]

**Beyond the Classical Model of the Inverse Faraday Effect**

Beyond these classical treatments, various quantum-theoretical models of the IFE have also been proposed.[40,43,47] Indeed, the original theoretical framework by Pershan *et al.* (1966) was itself quantum-based.[15] Subsequent studies have provided further insight into the quantum-mechanical response of materials under circularly polarized (CP) light excitation. For instance, Battiato *et al.* (2014) present a theory well-suited for ab-initio band-structure computations,[43] as later demonstrated by Berritta *et al.* (2016).[46] Their approach enables calculation of both spin and orbital contributions to the induced magnetic moment in materials with different magnetic orders. These findings may clarify certain experimental puzzles in all-optical switching (AOS). Specifically, for non-magnetic metals, their results suggest the spin contribution to the induced magnetization is an order of magnitude smaller than its orbital counterpart and arises solely from spin-orbit coupling.[46]

Additional insight was provided by Sinha-Roy *et al.* through time-dependent density-functional theory (TD-DFT) simulations of metallic nanospheres only a few nanometers in diameter.[41,42] The authors showed that a pronounced plasmonic resonance at the excitation wavelength is essential for producing a stable, stationary magnetic moment, which arises from light-driven, time-averaged circulating currents within the particle.[41] Similar findings had previously been reported with a semiclassical quantum-hydrodynamic model.[40] The atomic-scale resolution of TD-DFT further reveals quantum-confinement effects, such as Friedel-like oscillations in the current density,[41] and orbital-dependent contributions to the induced moment,[42] illustrated in Figure 2.

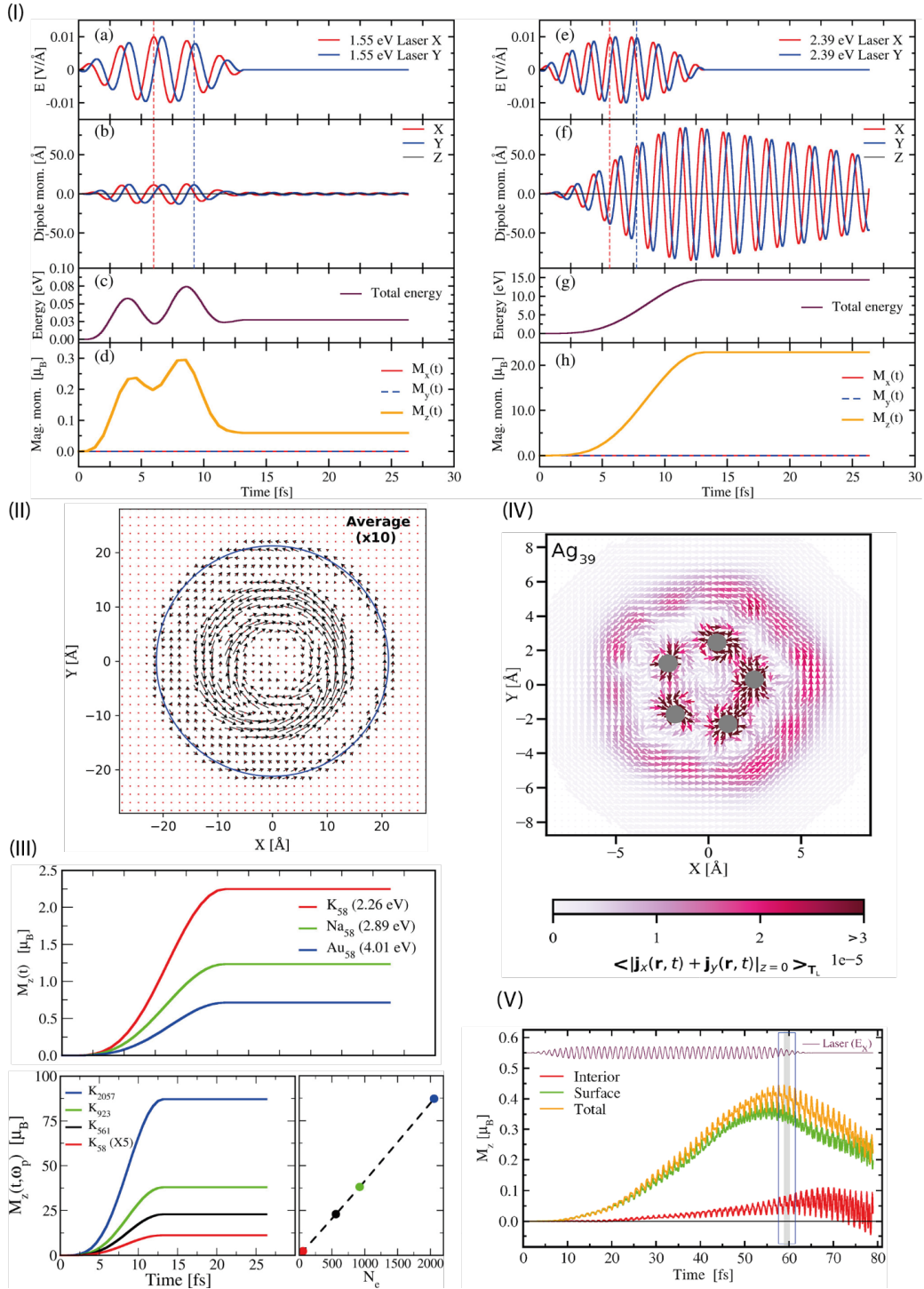

*Figure 2:* TD-DFT studies of the inverse Faraday effect in metallic nanoclusters. (I) Simulated jellium $K_{561}$ cluster under (left) off-resonant and (right) resonant excitation with a quasi-monochromatic, circularly polarized laser. (II) Time-averaged current-density map in $K_{561}$; the average is taken over one oscillation period of the induced dipole moment. (III) Top: $M_z$ component versus chemical composition for clusters containing the same total number of electrons, each excited at its localized surface-plasmon (LSP) frequency. Bottom: $M_z$ versus electron count for otherwise identical clusters. (IV) Atomic-scale results for an Ag cluster. Top: time-averaged current-density distribution (Ag atoms shown as gray circles). Bottom: separate contributions from the core and surface regions to the total magnetization $M_z$. Panels (I-III) are adapted from reference [41]. Panels (IV-V) are adapted from reference [42].

However, due to the high computational cost, these methods are limited to very small systems—just a few nanometers in size. In contrast, well-established finite-element electromagnetic simulations combined with a classical description of the phenomenon allow for the exploration of larger systems. As we will discuss later in this review, this approach is already sufficient for investigating and fostering meaningful discussion around the IFE.

**Drift Photocurrents and Their Role in the Inverse Faraday Effect**

Despite its apparent simplicity, Hertel's classical theory has created some confusion in the community. This theory closely links the inverse Faraday effect to the generation of drift photocurrents, yet the different contributions to these currents—and their relationship to the induced magnetization—remain somewhat misunderstood.

Adhering to a classical description that neglects spin moments, Hertel and Fähnle analytically examined a gold film illuminated by a circularly polarized Gaussian beam.[13] Their analysis identified two distinct contributions to the total induced magnetic moment, arising from two types of circulating electron motion: one microscopic and one macroscopic. The first (microscopic) contribution comes from the local gyroscopic motion of electrons, which underpins the magnetization $\vec{M}$ described earlier. The second (macroscopic) contribution stems from "drift currents," representing electronic motion on slower timescales. This macroscopic drift is driven by field gradients that, over several optical cycles, cause the electrons to deviate from perfectly harmonic circular motion (Equation 14). Although the second contribution is often smaller, the authors emphasize that both contributions must be considered to avoid overestimating the IFE, as they can have opposing signs.

A key point that is frequently overlooked is that the microscopic gyration of electrons can also be expressed in terms of its time-averaged value, corresponding to the magnetization currents previously described:

$$\vec{J_m} = \vec{\nabla} \times \vec{M} \quad (14)$$

Calculating the magnetic moment by integrating $\vec{M}$ over a structure or by applying the Biot–Savart law with $\vec{J_m}$ both describe the same physical contribution. As for the macroscopic drift currents in Hertel's 2015 work, it can be verified—within the same analytically resolved context—that the ponderomotive currents $\vec{\Gamma}$ match this description. Thus, by expressing both contributions as time-averaged currents, the inverse Faraday effect can be fully captured by the total drift photocurrents $\vec{J_d}$ described earlier (Equation 13).[48]

Building on this classical framework and computing circular charge trajectories, Nadarajah and Sheldon estimated the magnetic moment generated by the inverse Faraday effect in spherical Au nanoparticles (diameter ~100 nm), as well as the current density within the structures (Figure 3 (I)).[39] Their separation into two contributions reflects how they interpret the IFE in these spherical nanoparticles. They observe that the magnetic moment arising from macroscopic drift is about three orders of magnitude lower than that from micro-orbits, aligning with Hertel and Fähnle's simpler prediction. Notably, in the case of spherical nanoparticles, both contributions have the same sign, indicating that optical field gradients profoundly influence how macroscopic currents flow in a given structure.

Because of its straightforward implementation and relatively low computational demands, this description is well-suited to optimization approaches. For instance, Yang *et al*. employed a genetic algorithm to optimize a gold nanoantenna's structure by maximizing drift currents, proposing a design theoretically capable of producing intense magnetic fields in the tesla range under high excitation power (Figure 3 (II)).[45]

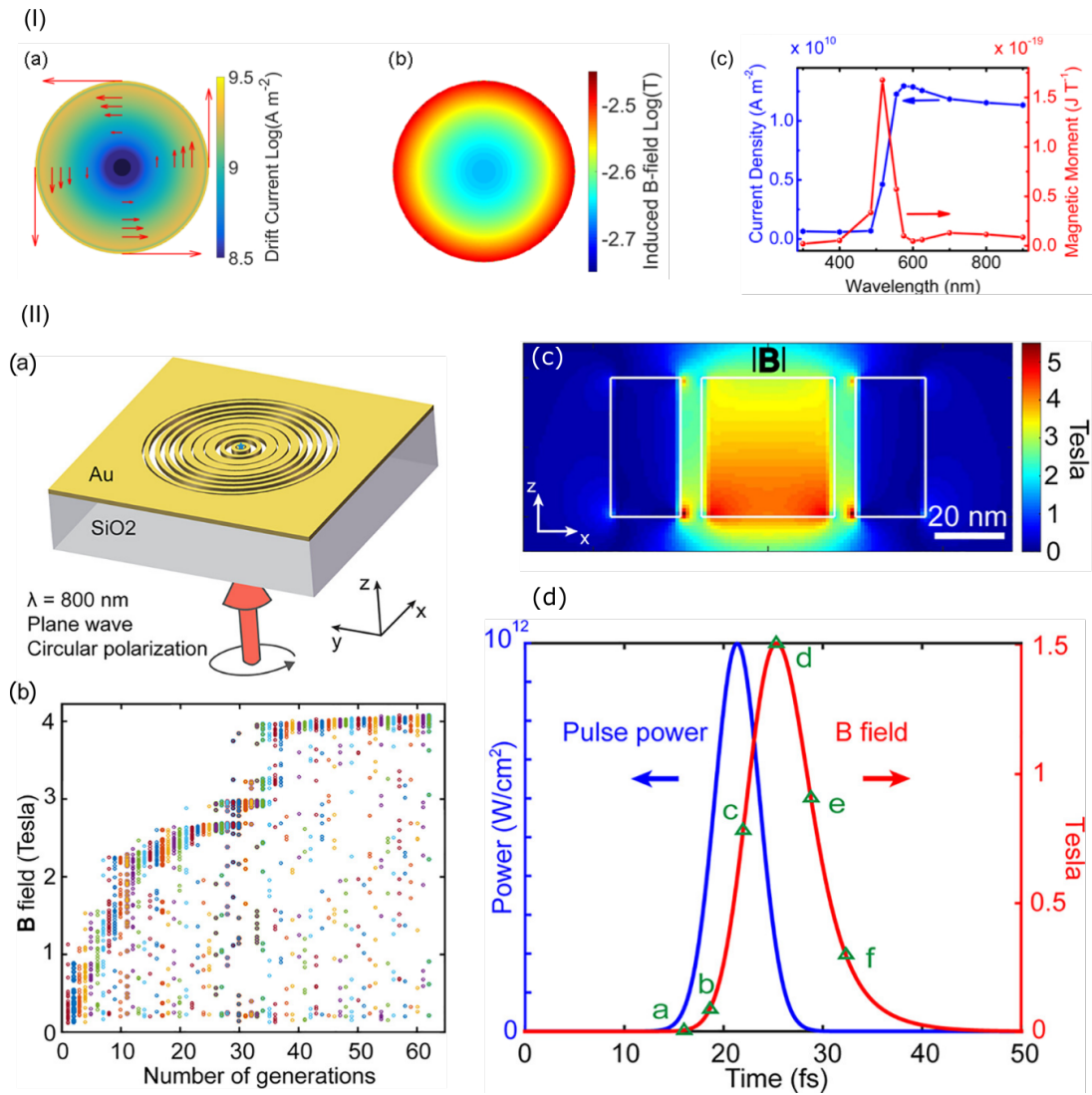

*Figure 3: Finite-element simulations of the IFE in plasmonic nanostructures.* *(I) Au nanosphere (100 nm diameter) excited at ≈517 nm with a $10^{11}$ W cm⁻² intensity, left-handed circularly polarized beam. (a) Calculated macroscopic drift-current density; (b) resulting static magnetic field; (c) wavelength dependence of the induced field, showing a clear resonance. Adapted from ref [39]. (II) Genetic-algorithm optimization of a plasmonic antenna that yields Tesla-scale magnetic fields under resonant pulsed excitation. (a) Optimized geometry; (b) selection map of candidate structures ranked by peak field strength; (c) spatial profile of the B-field at the structure center (star in panel a); (d) temporal evolution of the optical pulse (blue) and the induced magnetic field (red). Adapted from ref [45].*

## Plasmonic Pathways for Tailoring the Inverse Faraday Effect

As mentioned earlier, a simple classical framework suffices for numerical investigations that extend beyond merely calculating a system's typical response.[16,49–52] Indeed, Hertel's theory—especially when combined with optimization approaches—has shown that unconventional IFE responses can be achieved in specific plasmonic nanostructures. For example, it has been demonstrated that the IFE can be triggered by linearly polarized light in gold nanorods, which challenges the simplest definitions of the effect.[49] Moreover, the possibility of a *chiral* IFE—manifesting only for one helicity of the excitation—and even a *reversed* IFE, where the sign of the induced magnetic response runs counter to conventional expectations, has also been reported (see Figure 4).[16,50]

These properties arise from the way plasmonic nanoantennas manipulate the local optical field, particularly its polarization. One way to understand this manipulation is by studying the spin density $\vec{s}$ (Equation 15) of light near a plasmonic nanostructure—an inherently vector quantity whose components can adopt positive or negative values, corresponding to right- or left-handed elliptical polarizations, respectively. In a Cartesian coordinate system where the positive z-axis aligns with the direction of light propagation, a positive spin density along the propagation direction of light indicates right-handed helicity, a negative spin density denotes left-handed helicity, and zero corresponds to linear polarization. Although the far-field spin density typically ranges from -1 to +1 (with -1 signifying left-circular polarization and +1 right-circular polarization), in the near field this quantity—normalized by the incident intensity $|E_0|^2$—can significantly exceed those limits due to strong field enhancements. This phenomenon, referred to as "super-circular" light, draws an analogy to *super-chiral* light.[49,53]

$$\vec{s} = \frac{1}{|E_0|^2} Im(\vec{E^*} \times \vec{E}) \qquad (15)$$

Hence, examining the near-field spin density is particularly well-suited for assessing the generation of drift currents in a plasmonic structure, as increases in spin density directly correlate with the intensity of these currents (Equation 12). In fact, recent investigations indicate that by tailoring the spin-density distribution around such structures, it is theoretically possible to control the circulation of total drift photocurrents. This, in turn, can produce unconventional IFE responses that may hold promise for electronic applications, especially where on-demand switching is required.[16,49,50,54]

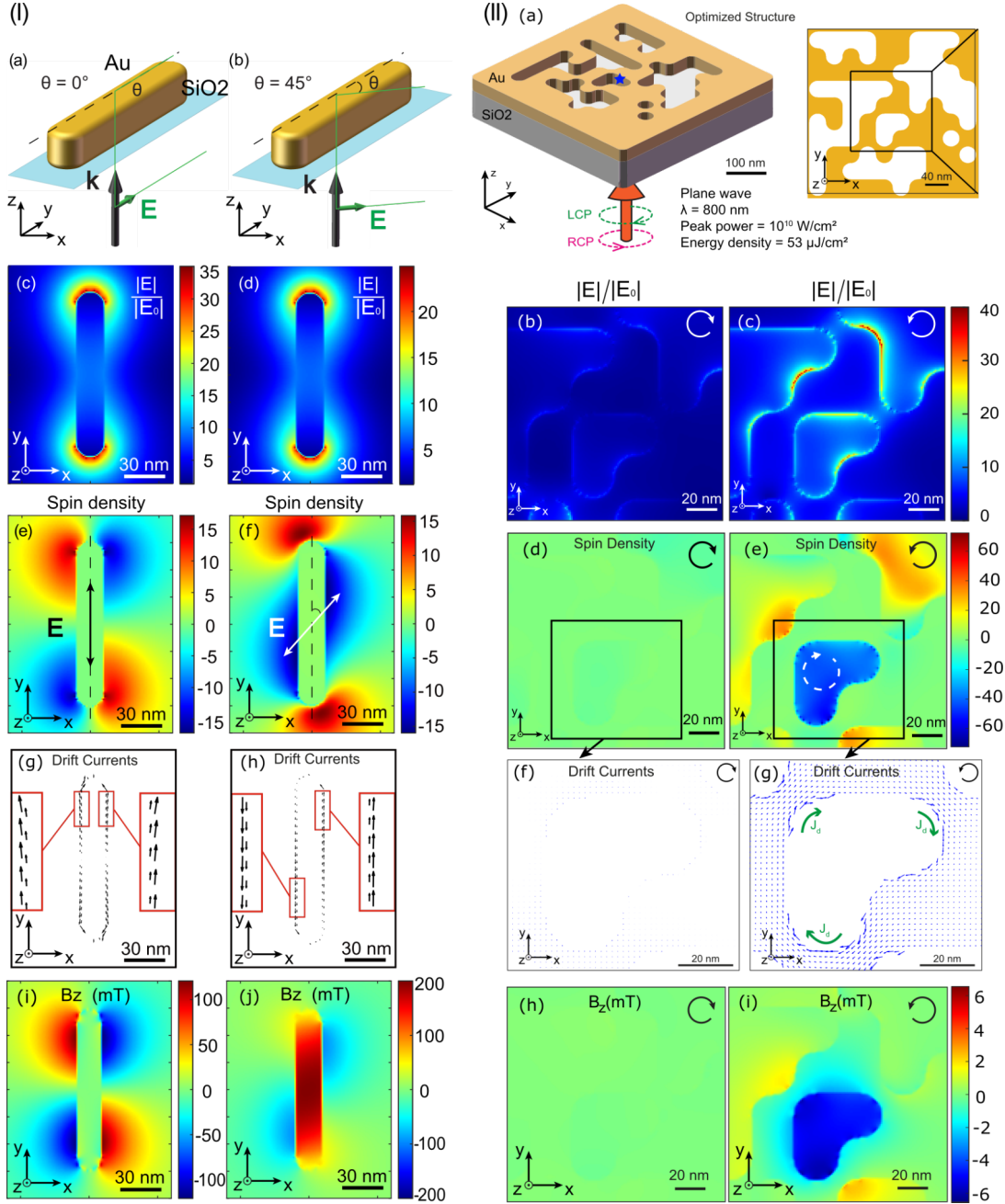

*Figure 4*: Role of Spin-density in the engineering of the inverse Faraday effect in plasmonic nanostructures. (I) Gold nanorod excited with linearly polarized light. Two configurations are compared: (a) polarization aligned with the rod's long axis and (b) polarization tilted 45° to that axis. For each case we plot the optical electric field (c,d), spin density (e,f), drift-photocurrent density (g,h), and the $B_z$ component of the IFE-induced magnetic field (i,j). The results demonstrate that a linearly polarized beam can drive an inverse Faraday effect when the spin density is redistributed by the plasmonic mode. Adapted from ref [49]. (II) Plasmonic antenna optimized by a genetic algorithm. (a) Final geometry. Under left- (LCP) and right-handed circular polarization (RCP) we show: optical electric field (b,c), spin density (d,e), drift-photocurrent density (f,g), and $B_z$ map (h,i). The design yields a **reversed** IFE (magnetization sign opposite to the usual helicity rule) that is also **chiral**, appearing only for one helicity. These studies confirm that plasmonic structures, by locally shaping the spin density, allow fine tailoring of the inverse Faraday effect. Adapted from ref [16].

# Probing the Inverse Faraday Effect: A Survey of Experimental Approaches

Experimental evidence of optically induced magnetization in plasmonic structures has been reported through various techniques. In this section, we present an overview of some of these characterizations.

In the early 2000s, Smolyaninov *et al.* investigated an array of nanoholes in a non-magnetic metallic sample using magnetic force microscopy (see Figure 5(I)). By employing a highly sensitive probe, they observed a magnetic contrast emerging only when the sample was illuminated. The observed pattern closely followed the sample's topography, suggesting a link between the detected magnetic interactions and the plasmonic features (i.e., the nanoholes).[55] Although the authors did not explicitly attribute their findings to the inverse Faraday effect—referring instead to "surface vacuum diamagnetism" and "surface-plasmon-induced magnetization"—their results align with our current understanding of the IFE. These observations encouraged subsequent attempts to measure the effect using magnetometry and near-field techniques.

In 2014, Moocarne *et al.* examined the magneto-optical activity in an aqueous colloidal solution of spherical gold nanoparticles (Figure 5(III)). By measuring the transmission spectra under external magnetic fields of varying intensities (with a fixed optical excitation intensity of 1 W/cm$^2$), they identified different features in the transmission spectra—especially near the plasmon resonance—across three regimes: low (0–0.5 T), intermediate (<1 T), and high (>1 T) magnetic fields. They proposed a theoretical model that accurately reproduces the observed spectra, suggesting a transition between linear and nonlinear behavior, wherein the nonlinear behavior under high magnetic fields can be described using a classical IFE framework.[34] This study provides valuable insight into how the system's properties can change substantially across different magnetic field regimes.

In 2022, Sheldon *et al.* investigated how the IFE affects plasmon damping in achiral gold nanodisks (Figure 5(II)), employing a Raman thermometry technique. Their experiments revealed increased backscattering and absorption under circularly polarized light as compared to linearly polarized light, which they attributed to reduced plasmon damping stemming from the IFE. The decreased damping intensified photothermal heating, whose local temperature they measured using Raman-based thermometry. Moreover, upon applying an external magnetic field, the photothermal heating remained lower for one helicity of light over a range of excitation intensities. The authors interpreted this observation as a result of competing Lorentz forces and IFE-induced microscopic electron orbits; when the IFE-induced magnetization is antiparallel to the applied field, the plasmon damping increases.[56]

Although these investigations do not provide direct quantitative measurements of the inverse Faraday effect, they demonstrate how this light–matter interaction in metals can both induce magnetization and actively alter intrinsic material properties.

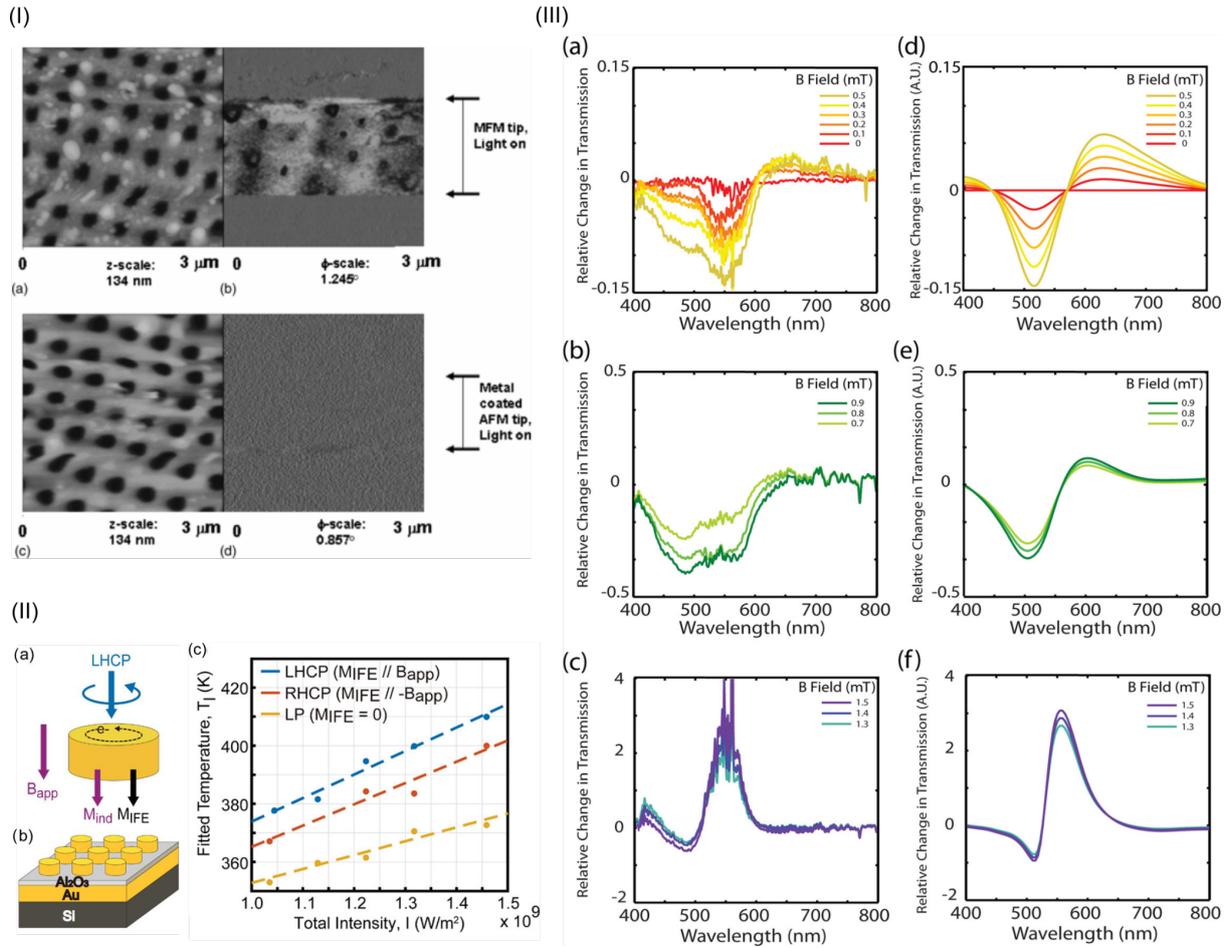

*Figure 5*: Indirect probes of the IFE in plasmonic nanostructures. (I) Magnetic-force microscopy (MFM) of a nano-hole array under 488 nm illumination. Phase images recorded with a magnetic tip (a) and a non-magnetic tip (c) are compared with the corresponding topography maps (b,d). Only the central region was exposed to light. A magnetic contrast appears solely under illumination with the magnetic probe, indicating that the light-induced interaction is linked to the nano-hole pattern. Adapted from ref [55]. (II) Photothermal thermometry on a gold nanodisk array in a 0.2 T static field. Experimental layout is sketched in (a,b). Panel (c) shows the fitted temperature rise versus incident intensity for linear and circular polarizations. Higher heating for circular polarization indicates that IFE-induced magnetization reduces plasmon damping, concentrating optical energy at hot spots; the effect is strongest when the IFE magnetization is parallel to the external field. Adapted from ref [56]. (III) Magneto-optical transmission of an 80 nm-diameter Au nanoparticle colloid excited with right-handed circular polarization (1 W cm$^{-2}$). Experimental spectra at low, intermediate, and high external fields are shown in (a–c); theoretical fits based on a linear Drude model (d), a nonlinear IFE model (f), and their combination (e) capture the transition between regimes. Adapted from ref [34].

While the aforementioned experiments utilized continuous-wave excitation, the growing interest in ultrafast magnetization dynamics has prompted researchers to investigate the inverse Faraday effect under pulsed laser excitation. This has enabled the first quantitative characterizations of the phenomenon and offered a clearer understanding of its temporal evolution.

# Quantitatively measuring the IFE via pump-probe experiments

The first experimental quantification of the inverse Faraday effect in non-magnetic metals was reported in 2019 by Cheng and Sheldon, who examined a colloidal system of 100 nm spherical gold nanoparticles (AuNPs) excited by an intense pulsed laser near their plasmonic resonance.[57] Utilizing a pump–probe setup, they obtained time-resolved measurements of the Faraday rotation—stemming from the conventional Faraday effect—of a linearly polarized probe beam after it interacted with the IFE-magnetized AuNPs (Figure 6(I)). In this straightforward detection scheme, and by monitoring the probe polarization, they demonstrated that their colloidal AuNP solution was magnetized by the pump beam on sub-picosecond timescales (Figure 6(I)). Furthermore, consistent with the IFE, the rotation exhibited opposite signs for left- and right-handed circular polarization, while scaling linearly with the peak pump intensity. By using the IFE-related Verdet constant, the measured rotation can be related to a magnetization (Figure 6 (I)).

These experiments paved the way for others to start studying IFE in different systems of non-magnetic metals, such as thin films,[36] or nanodisks,[58] or – more recently – plasmonic-ferromagnetic heterostructures.[37]
These recent experimental results show that plasmonic structures allow an enhancement of the IFE by at least one order of magnitude as compared to an excitation with no resonance effects.

The use of other parallel experimental methods and characterizations has allowed for insight into how IFE might depend on the material properties. A good example is the recent work of Gonzalez-Alcalde *et al*.[58] The authors use both static and time-resolved thermo-transmission measurements to determine the LSPR of the studied gold nanodisks, together with the time-resolved IFE measurements using Faraday rotation (Figure 6(II)). This allows them to correlate the two signals. As expected, they observe that the structures present an enhanced IFE (up to one order of magnitude compared to a bare gold film) when the LSPR is closer to the pump excitation.

In another example, the authors in reference [36] use Kerr rotation to perform time-resolved IFE measurements on thin film samples of different transition metals. Their study aims to understand which material properties play a role in IFE. Thus, they characterize their samples using X-ray diffraction, AFM, resistivity, absorptance, and ellipticity measurements to assess the crystal structure, electrical conductivity and optical properties. Interestingly, the results indicate that despite significant differences in these properties, the IFE is of similar magnitude for all of the thin metal films. They attribute the slightly higher values to a stronger spin-orbit coupling in the materials together with an interband transition near the excitation which are hypothesized to induce additionally spin and orbital moment contributions. Nevertheless, the effect remains small in magnitude and hard to compare to the existing theories.

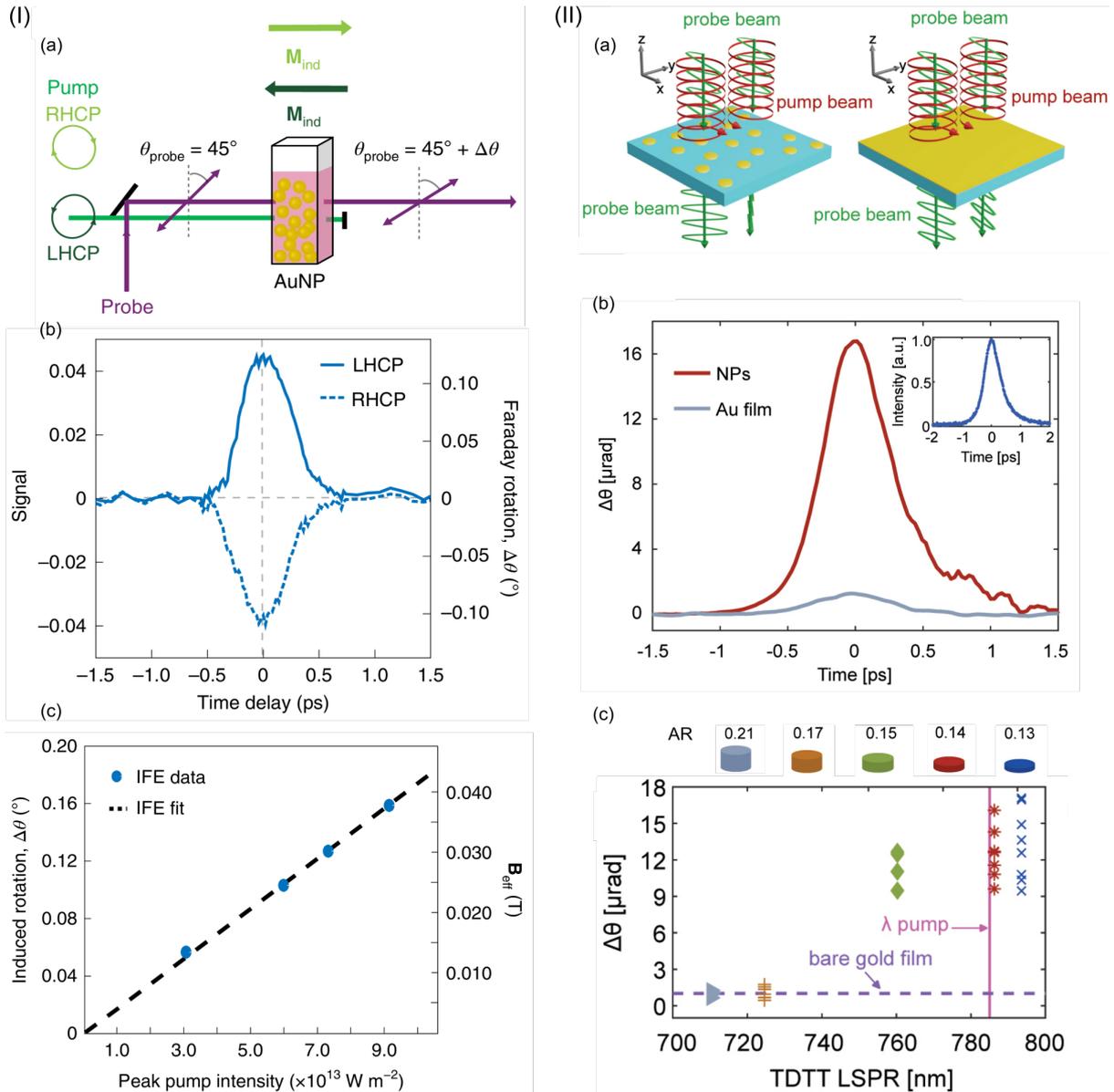

*Figure 6*: Time-resolved pump–probe studies of the inverse Faraday effect (IFE). (I) (a) Schematic of the pump–probe setup. A circularly polarized pump pulse magnetizes the nanoparticles via the IFE; the resulting magnetization rotates the linear polarization of the transmitted probe through the conventional Faraday effect. (b) The photo-induced rotation reverses sign when the pump helicity changes from left-handed to right-handed, confirming a purely IFE origin. (c) IFE rotation scales linearly with pump power. Using the IFE-related Verdet constant, the rotation angle can be converted to an effective magnetic field, providing a direct measure of the IFE strength. Adapted from ref [57]. (II) Plasmonic enhancement of the IFE in gold nanodisks. (a) Experimental scheme identical to (I), applied to gold nanodisks of various aspect ratios (ARs) and to a 20-nm-thick gold film. (b) The IFE signal from nanodisks is at least one order of magnitude larger than that from the continuous film. Inset: autocorrelation trace of the pump pulse, showing that the IFE dynamics follow the pump envelope for both samples. (c) Peak IFE rotation versus localized surface-plasmon resonance (LSPR) wavelength for nanodisks with different ARs. Disks whose LSPR is closer to the pump wavelength exhibit the strongest enhancement. Adapted from ref [58].

## Bridging Theory and Experiment: Key Insights and Discussion

Designing experiments that can be directly benchmarked against theory is inherently challenging, and the inverse Faraday effect (IFE) in nanoplasmonic systems is no exception. To date, several theoretical frameworks—both classical and quantum—have been proposed, with each underpinning a specific family of numerical methods. Because every simulation approach is optimized for a particular size or time scale, no single model spans the full range of experimentally relevant dimensions.

On the experimental side, most reports are qualitative; the few quantitative studies rely on ensemble measurements averaged over many nanostructures. Although the ultrafast dynamics observed experimentally agree with theoretical predictions—that is, the IFE-induced magnetization tracks the excitation envelope—the absolute magnitude of the effect remains difficult to reconcile with existing models.[36,39–41,56,58]

Tables 1 and 2 compile inverse Faraday effect data for gold systems obtained from experiment (Table 1) and theory (Table 2). Pump–probe measurements on various plasmonic nanostructures are mutually consistent, yielding similar orders of magnitude and showing the expected enhancement relative to continuous gold films. In contrast, theoretical estimates—derived from disparate models and numerical techniques—span several orders of magnitude, even for ostensibly comparable structures.

Whether these theoretical values can—or should—be cross-compared is not straightforward. For example, quantum TD-DFT simulations conserve total energy, so a continuous optical drive would lead to a magnetization that increases without bound.[41] By contrast, the FDTD study in ref [39] adopts a continuous excitation but reaches a steady-state IFE response after a sufficient number of time steps. Quantum-confinement effects offer another cautionary note: TD-DFT predicts pronounced Friedel-type current oscillations in small clusters,[41,42] whereas larger particles are dominated by surface currents and behave more classically. Finally, the optical response itself changes with scale; even at resonance, nano- and mesoscale plasmonic systems can exhibit markedly different field distributions and damping rates. For these reasons, it is not obvious that all results can be recast in universal units that allow a meaningful, one-to-one comparison across very different regimes.

The disparity is evident even for a seemingly straightforward system such as spherical gold nanoparticles. In 100 nm particles, the magnetization inferred from pump–probe experiments[57] exceeds the value predicted by FDTD simulations[39] by several orders of magnitude. This gap could reflect shortcomings in the theoretical model, in the experimental quantification, or in both. It may also signal that continuous-wave and pulsed excitation generate fundamentally different IFE responses, beyond the obvious difference in average power.

Rather than discouraging further work, this mismatch highlights the need for improved experimental methodologies. Current "magnetic" values are extracted from optical data through a conversion that relies on the Verdet constant; because the measurements do not probe magnetic fields directly, the constant must be estimated by combining a static Faraday-rotation measurement with a theoretical expression that introduces multiple assumptions.[43,57] Consequently, published IFE magnitudes are not purely experimental; they remain partly anchored to theory. Closing this loop—by developing techniques that measure the induced magnetization directly—will be essential for resolving the present discrepancy and for benchmarking future models.

*Table 1: Pump–probe measurements of the inverse Faraday effect in gold-based systems*

| Ref | System | Excitation | Measured IFE rotation [μrad.m$^2$/TW] | IFE rotation per atom [μrad.m$^2$/TW] |
|---|---|---|---|---|
| Cheng *et al*. 2020 [57] | colloidal Au NPs (100 nm diameter) | center wavelength: 515 nm<br><br>estimated pulse width: 370 fs | 30 | 9.8x10$^{-7}$ |
| Ortiz *et al*. 2023 [36] | Au thin film (20 nm thickness) | center wavelength: 783 nm<br><br>estimated pulse width: 330 fs | 1.5 | 1.3x10$^{-9}$ |
| Gonzales-Alcalde *et al*. 2024 [58] | Au nanodisk (190nm diameter, 24 nm thickness) | center wavelength: 785 nm<br><br>estimated pulse width: 430 fs | 24 | 6x10$^{-7}$ |

*Table 2: Reported magnitudes of the inverse Faraday effect in gold structures: theory vs experiment*

| Ref | System | Model/method | Magnetic moment per atom [μB] | Power density normalized magnetic moment [μB·m$^2$/TW] | Energy density normalized [μB·m$^2$/(TW·fs)] |
|---|---|---|---|---|---|
| Nadarajah *et al*. 2017 | Au NP 100 nm | Classical model FDTD | 5.9x10$^{-4}$ | 5.9x10$^{-7}$ | – |
| [39] | diameter | simulations | | | |
| Hurst et al. 2018 [40] | Au, 1-6 nm diameter | Quantum hydro-dynamics (semi-classical) | 0.35 | 7.8x10$^{-5}$ | – |
| Sinha-Roy et al. 2020 [41] | Au cluster of 254 electrons (size < 6nm) | Quantum Jellium model TDDFT simulations | 8.6x10$^{-3}$ | 6.5x10$^{-4}$ | 4.9x10$^{-5}$ |
| Cheng et al. 2020 [57] | colloidal Au NP 100 nm diameter | Experimental quantification via Verdet constant | 0.95 | 1x10$^{-2}$ | 2.7x10$^{-5}$ |

Additional open questions include the role of electron spin in the emerging magnetization,[36,44,46] and the recurrent observation that the wavelength of maximum IFE efficiency does not coincide with the plasmon-absorption peak.[37,46,57] Both issues connect to the broader debate on all-optical switching mechanisms and the possible involvement of electronic transitions and optical absorption. Another key point is the IFE-driven modification of a material's intrinsic electronic properties, recently reported in Ref [56]. Such concurrent changes in electronic and optical response are largely ignored in current models and may help explain the disparity between theoretical and experimental magnitudes.

Although simulations are invaluable for guiding experiments, they seldom capture the full complexity of nanostructure–light interactions. Experimental quantification of the IFE must disentangle magnetization from concurrent phenomena such as transient refractive-index changes, altered plasmon damping, electronic heating, and thermal diffusion.[36–38,42,56,57,59] Refining computational tools to incorporate these coupled effects could close the theory–experiment gap.

On the experimental side, progress hinges on developing robust, high-sensitivity probes of optically induced magnetic fields in plasmonic nanostructures. The community still lacks a clear picture of the attainable field strengths and spatial profiles. New methods must operate deep in the nonlinear regime where the IFE emerges, while offering flexibility in excitation wavelength and nanostructure geometry. Improved experimental characterization will, in turn, feed back into theory, fostering convergence toward a unified description of the inverse Faraday effect at the nanoscale.

## Conclusion and perspectives

Interest in the inverse Faraday effect (IFE) has surged over the past decade, driven by rapid advances in light-induced magnetism and nanoplasmonics. In principle, the IFE offers a route to ultrafast, nanoscale control of magnetic order. Several theoretical frameworks—both classical and quantum—capture different facets of the phenomenon, yet a unified description remains elusive. Even so, these models have been leveraged in numerical simulations that shed light on key ingredients: local spin density, electron density, field gradients, confinement effects, and more. When combined with the design freedom of plasmonic nanostructures, simulations predict that IFE can be deliberately tailored, giving rise to unexpected behaviors such as linear-polarization-driven IFE, chiral or "one-handed" IFE, and even helicity-reversed IFE.

Experimentally, IFE has been invoked to explain a variety of observations, including nonlinear magneto-optical responses, all-optical switching, and transient changes in plasmon damping. More recently, semi-quantitative pump–probe studies have begun to report IFE magnitudes in plasmonic systems. These time-resolved measurements confirm the anticipated sub-picosecond dynamics and show enhanced signals near plasmonic resonances. However, direct comparison with theory is still problematic. A fully experimental, magnetically sensitive probe that yields absolute IFE amplitudes has yet to be demonstrated, and no study has mapped the static magnetic-field profile with nanoscale resolution.

The next stage of progress will hinge on improved experimental platforms capable of testing the increasingly complex nanostructures proposed in theory. High-sensitivity, high-resolution tools—ideally operating in the nonlinear regime where IFE manifests—will be essential for closing the gap between prediction and observation and for realizing the full potential of ultrafast magnetism at the nanoscale.

(1) Kimel, A. V.; Li, M. Writing Magnetic Memory with Ultrashort Light Pulses. *Nat. Rev. Mater.* **2019**, *4* (3), 189–200. https://doi.org/10.1038/s41578-019-0086-3.
(2) Scheid, P.; Remy, Q.; Lebègue, S.; Malinowski, G.; Mangin, S. Light Induced Ultrafast Magnetization Dynamics in Metallic Compounds. *J. Magn. Magn. Mater.* **2022**, *560*, 169596. https://doi.org/10.1016/j.jmmm.2022.169596.
(3) Bossini, D.; Belotelov, V. I.; Zvezdin, A. K.; Kalish, A. N.; Kimel, A. V. Magnetoplasmonics and Femtosecond Optomagnetism at the Nanoscale. *ACS Publications*. https://doi.org/10.1021/acsphotonics.6b00107.
(4) Armelles, G.; Cebollada, A.; García-Martín, A.; González, M. U. Magnetoplasmonics: Combining Magnetic and Plasmonic Functionalities. *Adv. Opt. Mater.* **2013**, *1* (1), 10–35. https://doi.org/10.1002/adom.201200011.
(5) Beaurepaire, E.; Merle, J.-C.; Daunois, A.; Bigot, J.-Y. Ultrafast Spin Dynamics in Ferromagnetic Nickel. *Phys. Rev. Lett.* **1996**, *76* (22), 4250. https://doi.org/10.1103/PhysRevLett.76.4250.
(6) Kimel, A. V.; Kirilyuk, A.; Usachev, P. A.; Pisarev, R. V.; Balbashov, A. M.; Rasing, T. Ultrafast Non-Thermal Control of Magnetization by Instantaneous Photomagnetic Pulses. *Nature* **2005**, *435* (7042), 655–657. https://doi.org/10.1038/nature03564.
(7) Stanciu, C. D.; Hansteen, F.; Kimel, A. V.; Kirilyuk, A.; Tsukamoto, A.; Itoh, A.; Rasing, T. All-Optical Magnetic Recording with Circularly Polarized Light. *Phys. Rev. Lett.* **2007**, *99* (4), 047601. https://doi.org/10.1103/PhysRevLett.99.047601.
(8) Lambert, C.-H.; Mangin, S.; Varaprasad, B. S. D. C. S.; Takahashi, Y. K.; Hehn, M.; Cinchetti, M.; Malinowski, G.; Hono, K.; Fainman, Y.; Aeschlimann, M.; Fullerton, E. E. All-Optical Control of Ferromagnetic Thin Films and Nanostructures. *Science* **2014**. https://doi.org/10.1126/science.1253493.
(9) Khorsand, A. R.; Savoini, M.; Kirilyuk, A.; Kimel, A. V.; Tsukamoto, A.; Itoh, A.; Rasing, T. Role of Magnetic Circular Dichroism in All-Optical Magnetic Recording. *Phys. Rev. Lett.* **2012**, *108* (12), 127205. https://doi.org/10.1103/PhysRevLett.108.127205.
(10) Kirilyuk, A.; Kimel, A. V.; Rasing, T. Ultrafast Optical Manipulation of Magnetic Order. *Rev. Mod. Phys.* **2010**, *82* (3), 2731. https://doi.org/10.1103/RevModPhys.82.2731.
(11) Hertel, R. Theory of the Inverse Faraday Effect in Metals. *J. Magn. Magn. Mater.* **2006**, *303* (1), L1–L4. https://doi.org/10.1016/j.jmmm.2005.10.225.
(12) Zhang, H.-L.; Wang, Y.-Z.; Chen, X.-J. A Simple Explanation for the Inverse Faraday Effect in Metals. *J. Magn. Magn. Mater.* **2009**, *321* (24), L73–L74. https://doi.org/10.1016/j.jmmm.2009.08.023.
(13) Hertel, R.; Fähnle, M. Macroscopic Drift Current in the Inverse Faraday Effect. *Phys. Rev. B* **2015**, *91* (2), 020411. https://doi.org/10.1103/PhysRevB.91.020411.
(14) Ziel, J. P. van der; Pershan, P. S.; Malmstrom, L. D. Optically-Induced Magnetization Resulting from the Inverse Faraday Effect. *Phys. Rev. Lett.* **1965**, *15* (5), 190. https://doi.org/10.1103/PhysRevLett.15.190.
(15) Pershan, P. S.; van der Ziel, J. P.; Malmstrom, L. D. Theoretical Discussion of the Inverse Faraday Effect, Raman Scattering, and Related Phenomena. *Phys. Rev.* **1966**, *143* (2), 574–583. https://doi.org/10.1103/PhysRev.143.574.
(16) Mou, Y.; Yang, X.; Gallas, B.; Mivelle, M. A Reversed Inverse Faraday Effect. *Adv. Mater. Technol.* **2023**, *8* (21), 2300770. https://doi.org/10.1002/admt.202300770.
(17) Seung Joon Lee, †; Zhiqiang Guan, ‡; Hongxing Xu, ‡ and; Martin Moskovits*, †. *Surface-Enhanced Raman Spectroscopy and Nanogeometry: The Plasmonic Origin of SERS*. ACS Publications. https://doi.org/10.1021/jp077422g.
(18) Gwo, S.; Wang, C.-Y.; Chen, H.-Y.; Lin, M.-H.; Sun, L.; Li, X.; Chen, W.-L.; Chang, Y.-M.; Ahn, H. *Plasmonic Metasurfaces for Nonlinear Optics and Quantitative SERS*. ACS Publications. https://doi.org/10.1021/acsphotonics.6b00104.
(19) Curto, A. G.; Volpe, G.; Taminiau, T. H.; Kreuzer, M. P.; Quidant, R.; Hulst, N. F. van. Unidirectional Emission of a Quantum Dot Coupled to a Nanoantenna. *Science* **2010**. https://doi.org/10.1126/science.1191922.
(20) Dmitriev, P. A.; Lassalle, E.; Ding, L.; Pan, Z.; Neo, D. C. J.; Valuckas, V.; Paniagua-Dominguez, R.; Yang, J. K. W.; Demir, H. V.; Kuznetsov, A. I. Hybrid Dielectric-Plasmonic Nanoantenna with Multiresonances for Subwavelength Photon Sources. *ACS Photonics* **2023**. https://doi.org/10.1021/acsphotonics.2c01332.

(21) Reynier, B.; Charron, E.; Markovic, O.; Gallas, B.; Ferrier, A.; Bidault, S.; Mivelle, M. Nearfield Control over Magnetic Light-Matter Interactions. *Light Sci. Appl.* **2025**, *14* (1), 1–9. https://doi.org/10.1038/s41377-025-01807-z.

(22) Martín-Hernández, R.; Grünewald, L.; Sánchez-Tejerina, L.; Plaja, L.; Jarque, E. C.; Hernández-García, C.; Mai, S. Optical Magnetic Field Enhancement Using Ultrafast Azimuthally Polarized Laser Beams and Tailored Metallic Nanoantennas. *Photonics Res.* **2024**, *12* (5), 1078–1092. https://doi.org/10.1364/PRJ.511916.

(23) Heras, A. de las; Bonafé, F. P.; Hernández-García, C.; Rubio, A.; Neufeld, O. Tunable Tesla-Scale Magnetic Attosecond Pulses through Ring-Current Gating. *J. Phys. Chem. Lett.* **2023**. https://doi.org/10.1021/acs.jpclett.3c02899.

(24) Juan, M. L.; Righini, M.; Quidant, R. Plasmon Nano-Optical Tweezers. *Nat. Photonics* **2011**, *5* (6), 349–356. https://doi.org/10.1038/nphoton.2011.56.

(25) Wang, D.; Guan, J.; Hu, J.; Bourgeois, M. R.; Odom, T. W. Manipulating Light–Matter Interactions in Plasmonic Nanoparticle Lattices. *Acc. Chem. Res.* **2019**. https://doi.org/10.1021/acs.accounts.9b00345.

(26) Estep, N. A.; Sounas, D. L.; Soric, J.; Alù, A. Magnetic-Free Non-Reciprocity and Isolation Based on Parametrically Modulated Coupled-Resonator Loops. *Nat. Phys.* **2014**, *10* (12), 923–927. https://doi.org/10.1038/nphys3134.

(27) Mohammadi, E.; Tittl, A.; Tsakmakidis, K. L.; Raziman, T. V.; Curto, A. G. Dual Nanoresonators for Ultrasensitive Chiral Detection. *ACS Photonics* **2021**. https://doi.org/10.1021/acsphotonics.1c00311.

(28) Collins, S. S. E.; Searles, E. K.; Tauzin, L. J.; Lou, M.; Bursi, L.; Liu, Y.; Song, J.; Flatebo, C.; Baiyasi, R.; Cai, Y.-Y.; Foerster, B.; Lian, T.; Nordlander, P.; Link, S.; Landes, C. F. Plasmon Energy Transfer in Hybrid Nanoantennas. *ACS Nano* **2020**. https://doi.org/10.1021/acsnano.0c08982.

(29) Baffou, G.; Quidant, R. Thermo-Plasmonics: Using Metallic Nanostructures as Nano-Sources of Heat. *Laser Photonics Rev.* **2013**, *7* (2), 171–187. https://doi.org/10.1002/lpor.201200003.

(30) Jain, P. K.; Xiao, Y.; Walsworth, R.; Cohen, A. E. Surface Plasmon Resonance Enhanced Magneto-Optics (SuPREMO): Faraday Rotation Enhancement in Gold-Coated Iron Oxide Nanocrystals. *Nano Lett.* **2009**, *9* (4), 1644–1650. https://doi.org/10.1021/nl900007k.

(31) Singh, N. D.; Moocarme, M.; Edelstein, B.; Punnoose, N.; Vuong, L. T. Anomalously-Large Photo-Induced Magnetic Response of Metallic Nanocolloids in Aqueous Solution Using a Solar Simulator. *Opt. Express* **2012**, *20* (17), 19214–19225. https://doi.org/10.1364/OE.20.019214.

(32) Belotelov, V. I.; Bezus, E. A.; Doskolovich, L. L.; Kalish, A. N.; Zvezdin, A. K. Inverse Faraday Effect in Plasmonic Heterostructures. *J. Phys. Conf. Ser.* **2010**, *200* (9), 092003. https://doi.org/10.1088/1742-6596/200/9/092003.

(33) Sepúlveda, B.; González-Díaz, J. B.; García-Martín, A.; Lechuga, L. M.; Armelles, G. Plasmon-Induced Magneto-Optical Activity in Nanosized Gold Disks. *Phys. Rev. Lett.* **2010**, *104* (14), 147401. https://doi.org/10.1103/PhysRevLett.104.147401.

(34) Moocarme, M.; Domínguez-Juárez, J. L.; Vuong, L. T. *Ultralow-Intensity Magneto-Optical and Mechanical Effects in Metal Nanocolloids*. ACS Publications. https://doi.org/10.1021/nl4039357.

(35) Pineider, F.; Campo, G.; Bonanni, V.; Fernández, C. de J.; Mattei, G.; Caneschi, A.; Gatteschi, D.; Sangregorio, C. *Circular Magnetoplasmonic Modes in Gold Nanoparticles*. ACS Publications. https://doi.org/10.1021/nl402394p.

(36) Ortiz, V. H.; Mishra, S. B.; Vuong, L.; Coh, S.; Wilson, R. B. Specular Inverse Faraday Effect in Transition Metals. *Phys. Rev. Mater.* **2023**, *7* (12), 125202. https://doi.org/10.1103/PhysRevMaterials.7.125202.

(37) Parchenko, S.; Hofhuis, K.; Larsson, A. Å.; Kapaklis, V.; Scagnoli, V.; Heyderman, L. J.; Kleibert, A. Plasmon-Enhanced Optical Control of Magnetism at the Nanoscale via the Inverse Faraday Effect. *Adv. Photonics Res.* **2024**, 2400083. https://doi.org/10.1002/adpr.202400083.

(38) Tsiatmas, A.; Atmatzakis, E.; Papasimakis, N.; Fedotov, V.; Luk'yanchuk, B.; Zheludev, N. I.; Abajo, F. J. G. de. Optical Generation of Intense Ultrashort Magnetic Pulses at the Nanoscale. *New J. Phys.* **2013**, *15* (11), 113035. https://doi.org/10.1088/1367-2630/15/11/113035.

(39) Nadarajah, A.; Sheldon, M. T. Optoelectronic Phenomena in Gold Metal Nanostructures Due to the Inverse Faraday Effect. *Opt. Express* **2017**, *25* (11), 12753–12764. https://doi.org/10.1364/OE.25.012753.

(40) Hurst, J.; Oppeneer, P. M.; Manfredi, G.; Hervieux, P.-A. Magnetic Moment Generation in Small Gold Nanoparticles via the Plasmonic Inverse Faraday Effect. *Phys. Rev. B* **2018**, *98* (13), 134439. https://doi.org/10.1103/PhysRevB.98.134439.

(41) Sinha-Roy, R.; Hurst, J.; Manfredi, G.; Hervieux, P.-A. Driving Orbital Magnetism in Metallic Nanoparticles through Circularly Polarized Light: A Real-Time TDDFT Study. *ACS Photonics* **2020**. https://doi.org/10.1021/acsphotonics.0c00462.

(42) Lian, D.; Yang, Y.; Manfredi, G.; Hervieux, P.-A.; Sinha-Roy, R. Orbital Magnetism through Inverse Faraday Effect in Metal Clusters. *Nanophotonics* **2024**, *13* (23), 4291–4302. https://doi.org/10.1515/nanoph-2024-0352.

(43) Battiato, M.; Barbalinardo, G.; Oppeneer, P. M. Quantum Theory of the Inverse Faraday Effect. *Phys. Rev. B* **2014**, *89* (1), 014413. https://doi.org/10.1103/PhysRevB.89.014413.

(44) Mishra, S. B.; Coh, S. Spin Contribution to the Inverse Faraday Effect of Nonmagnetic Metals. *Phys. Rev. B* **2023**, *107* (21), 214432. https://doi.org/10.1103/PhysRevB.107.214432.

(45) Yang, X.; Mou, Y.; Gallas, B.; Maitre, A.; Coolen, L.; Mivelle, M. Tesla-Range Femtosecond Pulses of Stationary Magnetic Field, Optically Generated at the Nanoscale in a Plasmonic Antenna. *ACS Nano* **2022**, *16* (1). https://doi.org/10.1021/acsnano.1c06922.

(46) Berritta, M.; Mondal, R.; Carva, K.; Oppeneer, P. M. *Ab Initio* Theory of Coherent Laser-Induced Magnetization in Metals. *Phys. Rev. Lett.* **2016**, *117* (13), 137203. https://doi.org/10.1103/PhysRevLett.117.137203.

(47) Popova, D.; Bringer, A.; Blügel, S. Theory of the Inverse Faraday Effect in View of Ultrafast Magnetization Experiments. *Phys. Rev. B* **2011**, *84* (21), 214421. https://doi.org/10.1103/PhysRevB.84.214421.

(48) Yang, X.; Hareau, C.; Gartside, J.; Mivelle, M. *Light-Driven Skyrmion Crystal Generation in Plasmonic Metasurfaces Through the Inverse Faraday Effect*. arXiv.org. https://arxiv.org/abs/2503.23800v1 (accessed 2025-04-17).

(49) Yang, X.; Mou, Y.; Zapata, R.; Reynier, B.; Gallas, B.; Mivelle, M. An Inverse Faraday Effect Generated by Linearly Polarized Light through a Plasmonic Nano-Antenna. *Nanophotonics* **2023**, *12* (4), 687–694. https://doi.org/10.1515/nanoph-2022-0488.

(50) Mou, Y.; Yang, X.; Gallas, B.; Mivelle, M. A Chiral Inverse Faraday Effect Mediated by an Inversely Designed Plasmonic Antenna. *Nanophotonics* **2023**, *12* (12), 2115–2120. https://doi.org/10.1515/nanoph-2022-0772.

(51) Xu, Z.-Y.; Xie, H. Photoinduced Currents and Inverse Faraday Effect in Graphene Quantum Dots. *Phys. Rev. B* **2024**, *110* (8), 085425. https://doi.org/10.1103/PhysRevB.110.085425.

(52) Ichiji, N.; Ishida, T.; Morichika, I.; Oue, D.; Tatsuma, T.; Ashihara, S. Designing Rotational Motion of Charge Densities on Plasmonic Nanostructures Excited by Circularly Polarized Light. *Nanophotonics* **2024**. https://doi.org/10.1515/nanoph-2024-0433.

(53) Tang, Y.; Cohen, A. E. Enhanced Enantioselectivity in Excitation of Chiral Molecules by Superchiral Light. *Science* **2011**. https://doi.org/10.1126/science.1202817.

(54) Mou, Y.; Yang, X.; Vega, M.; Zapata, R.; Gallas, B.; Bryche, J.-F.; Bouhelier, A.; Mivelle, M. Femtosecond Drift Photocurrents Generated by an Inversely Designed Plasmonic Antenna. *Nano Lett.* **2024**. https://doi.org/10.1021/acs.nanolett.4c00558.

(55) Smolyaninov, I. I.; Davis, C. C.; Smolyaninova, V. N.; Schaefer, D.; Elliott, J.; Zayats, A. V. Plasmon-Induced Magnetization of Metallic Nanostructures. *Phys. Rev. B* **2005**, *71* (3), 035425. https://doi.org/10.1103/PhysRevB.71.035425.

(56) Cheng, O. H.-C.; Zhao, B.; Brawley, Z.; Son, D. H.; Sheldon, M. T. Active Tuning of Plasmon Damping via Light Induced Magnetism. *Nano Lett.* **2022**. https://doi.org/10.1021/acs.nanolett.2c00571.

(57) Cheng, O. H.-C.; Son, D. H.; Sheldon, M. Light-Induced Magnetism in Plasmonic Gold Nanoparticles. *Nat. Photonics* **2020**, *14* (6), 365–368. https://doi.org/10.1038/s41566-020-0603-3.

(58) González-Alcalde, A. K.; Shi, X.; Ortiz, V. H.; Feng, J.; Wilson, R. B.; Vuong, L. T. Enhanced Inverse Faraday Effect and Time-Dependent Thermo-Transmission in Gold Nanodisks. *Nanophotonics* **2024**, *13* (11), 1993–2002. https://doi.org/10.1515/nanoph-2023-0777.

(59) Boeglin, C.; Beaurepaire, E.; Halté, V.; López-Flores, V.; Stamm, C.; Pontius, N.; Dürr, H. A.; Bigot, J.-Y. Distinguishing the Ultrafast Dynamics of Spin and Orbital Moments in Solids. *Nature* **2010**, *465* (7297), 458–461. https://doi.org/10.1038/nature09070.